\documentclass[aps,prd,groupedaddress,showpacs,twocolumn]{revtex4}

\usepackage{textcomp}
\usepackage{makeidx}
\usepackage{amsmath}
\usepackage{subfigure}
\usepackage{amssymb}
\usepackage{hyperref}
\usepackage{graphicx}
\usepackage{graphics}
\usepackage[utf8]{inputenc}
\usepackage{float}
\usepackage{color}
\usepackage{dsfont}

\begin{document}

\title{Extra packing of mass of anisotropic interiors induced by MGD}

\author{Cynthia Arias}
\email{cynthia.arias@yachaytech.edu.ec}
\affiliation{School of Physical Sciences \& Nanotechnology, Yachay Tech University, 100119
Urcuqu\'{i}, Ecuador}

\author{Francisco Tello-Ortiz}
\email{francisco.tello@ua.cl}
\affiliation{Departamento de F\'isica, Facultad de ciencias b\'asicas, Universidad de Antofagasta, Casilla 170, Antofagasta, Chile.}

\author{Ernesto Contreras}
\email{econtreras@usfq.edu.ec}
\affiliation{Departamento de F\'isica, Colegio de Ciencias e Ingenier\'ia, Universidad San Francisco de Quito, Quito, Ecuador}

%\date{\today}

\begin{abstract}
In this work we investigate the extra packing of mass within the framework of gravitational decoupling by means of Minimal Geometric Deformation approach. It is shown that, after a suitable set of the free parameters involved, the like--Tolman IV solution extended by Minimal Geometric Deformation not only acquire extra packing of mass but it corresponds to a stable configuration according to the adiabatic index criteria. Additionally, it is shown that the extra packing condition induce a lower bound on the compactness parameter of the seed isotropic solution and a stringent restriction on the decoupling parameter.
\end{abstract}
%\newpage 

%\pacs{}

\maketitle

\section{Introduction}

Seminal Tolman's work \cite{r1} about spherically symmetric and static fluid spheres driven by isotropic matter distribution 
%\i.e, equal radial and transverse pressures ($p_{r}=p_{\perp}=p$), 
is considered as a milestone regarding to the seeking of analytic solutions of Einstein field equations, which describe exciting structures such as neutron stars, white dwarfs etc. All these objects corresponding to the final stage in the life of stars provide us real laboratories (impossible to recreate on the Earth) to understand the behaviour in the strong gravitational field regime. Interestingly, 
some  features of these kind of configurations can be obtained from direct observations. For example, by measuring the surface gravitational red-shift we can infer 
the mass, $M$, and the radius, $R$,  of a star that constitute the so called compactness parameter, $u\equiv M/R$, which should be interpreted as quantification of how much mass can be packaged per unit radius before the gravitational collapse of the compact object takes place. In this respect, Buchdahl \cite{r2} determined that for any isotropic matter distribution the mass--radius ratio can not exceed the upper bound $M/R\leq 4/9$ that corresponds to a maximum gravitational surface red--shift $z_{s}=2$.  

However, as pointed out by Ruderman in his pioneering work \cite{r3}, celestial bodies are not necessarily made of isotropic matter but they could contain local anisotropies at least in certain very high density ranges $(\rho>10^{15} g/cm^{3})$, where the nuclear interactions must be treated relativistically. Similarly, the existence of super--Chandrasekhar white dwarfs that allow to explain overluminous type Ia supernovae demands that the Chandrasekar limit must be violated in
the presence of strong magnetic fields which may be treated as an anisotropic
fluid \cite{muk1,muk2,muk3,muk4,muk5,muk6,muk7}. Indeed, it is well known that a
magnetic field acting on a Fermi gas produces pressure anisotropy.

In this direction, the celebrated article by Bowers and Liang \cite{r4} laid the initial basis for the study of anisotropic structures within the framework of Einstein's general relativity (GR from now on). Regardingly, they found that the contributions coming from local anisotropies into the Tolman-Oppenheimer-Volkoff (TOV) equation \cite{r1,r5} is of Newtonian origin. 

The study of Buchdahl's ratio in different frameworks involving self--gravitating compact structures entails intriguing queries, such as: i) How is this limit modified in the presence of additional fields? ii) How much do corrections to Einstein--Hilbert's action influence mass--radius ratio? iii) Does the Buchdahl limit increase or decrease in the high dimension regime? Regardingly, these questions have been extensively studied
over the last two decades leading to interesting results  \cite{r6,r7,r8,r9,r10,r11,r12,r13,r14,r15,r16,r17,r18,r19,r20,r21,r22,r23,r24,r25,r26,r27,r28,r29,r30,r31,r32,r33,r34,r35,r36,r37,r38,r39,r40,r41,r42} and among all the mechanism behind
the modification of the Buchdahl's limit, the simplest one corresponds to introduce local anisotropies within the matter distribution \cite{r43} which lead to $u>4/9$, namely, stellar interiors with extra packing of mass. For example, in Ref. \cite{r42} it was shown that the introduction of a Kalb--Ramond (KR) field leads to a stable compact anisotropic configuration with extra packing of mass. It should be noticed that the KR field in \cite{r42} fill the whole space and as a consequence, the compact object is surrounded not by the Schwarzschild vacuum but by a background  corresponding to an exterior solution of Einstein equations sourced by a KR field.
In this respect, we may wonder if it is possible to induce extra packing of mass by introducing local anisotropies into the fluid distributions, without modifying
the Schwarzschild vacuum space--time.
Incidentally, the gravitational decoupling (GD) introduced in
\cite{r58,r62,r97} in the framework of GR results to be a natural manner to 
induce such modifications.\\
%been introduced to modify 
%To achieve this, we have employed a simple and powerful tool called gravitational decoupling by minimal geometric deformation (MGD) in the framework of compact structures describing stellar interiors \cite{r58,r62}.
As it is well known, GD is a generalization of
the Minimal Geometric Deformation (MGD)
%In principle this methodology was 
employed in the context of Randall--Sundrum brane--world  \cite{r46,new1,r53,r56,r47,r48,r49,r50,r51,r52,r54,r55,r57,r59} 
%and then spread to investigate new classes of black holes solutions, by deforming the well--known Schwarzschild space--time
%\cite{r53,r56}. The earlier applications of this approach were mainly developed on the stage of the brane--world \cite{r47,r48,r49,r50,r51,r52,r54,r55,r57,r59}, black hole acoustic \cite{r60} and GUP Hawking fermions \cite{r61}. After that, this scheme was translated into the GR arena to extent isotropic fluid distributions satisfying Einstein field equations to an anisotropic domains \cite{r58,r62}.
and as we shall review
in the next section, the methodology contains two main ingredients: i) two sources, $\bar{T}_{\mu\nu}$ (that usually corresponds to a perfect fluid) and
$\theta_{\mu\nu}$
which only interact gravitationally  and ii) a minimal geometric deformation introduced in the $g^{rr}$ component of the metric which allows to decouple the system into two set of equations, one for each source. It is worth mentioning that in many applications of the GD,
$\bar{T}_{\mu\nu}$
is the source of a well known isotropic interior solution, so the
effect of $\theta_{\mu\nu}$ is to introduce a local anisotropy in the system and consequently, it is said that GD leads to an extension of isotropic solutions to anisotropic domains.
In this case, either
the geometric deformation and
the components of $\theta_{\mu\nu}$ remain unknown and the main goal 
is to provide suitable extra constraints which allow to find them. In this regard, a wide range of possibilities have been proposed, among which are: i) the so--called mimic constraint procedure \cite{r62,r60,r61,r63,r64,r65,r66,r67}, ii) the imposition of an adequate decoupling function $f(r)$ meeting all the physical and mathematical requirements \cite{r68,r69} or iii) some anisotropy mechanism \cite{r70,r71}.\\ 
%Depending on the mechanism considered to close the $\theta$--sector the magnitude and sign of the dimensionless coupling constant $\alpha$ is determined in order to preserve a well defined anisotropy factor $\Delta\equiv p_{\perp}-p_{r}$ throughout the stellar medium.\\
Given the versatility of GD by MGD,
its application to deal with a variety of situations has grown considerably 
during the last two years
\cite{r72,r73,r74,r75,r76,r77,r78,r79,r80,r81,r82,r83,r84,r85,r86,r87,r88,r89,r90,r91,r92,r93,r94, roldao1, roldao2,r95,r97}.
%, what is more the inverse problem and the extended case were developed in \cite{r96,r97}, respectively.
However, although the modification
of the Buchdahl limit in the framework of GD  is straightforward, surprisingly it has not been considered yet to investigate the extra packing of mass of interior stellar.
%the framework of gravitational decoupling by means of MGD approach
For this reason, the main goal of this
work is to employ the B\"{o}mer and Harko methodology in \cite{r43}  to study the Buchdahl limit
for anisotropic compact stars 
but assuming that
%where 
the anisotropy is encoded in the $\theta$-sector induced by the GD approach. 
%This procedure allows  us to constraint the values of the decoupling parameter, $\alpha$, in order to describe a well behaved interior solution. What is more, the values obtained for $\alpha$ determine the size of the mass--radius ratio. As an example to illustrate the scheme we use the deformed Tolman IV solution already obtained in \cite{r62} employing the mimic constraint grasp. It is shown that $\alpha$ can not be neither arbitrarily large or negative in order to describe a well posed stellar interior medium with extra packing mass.
%It is worth mentioning that this is the first time that modifications on the well--known Buchdahl's limit are addressed in the context of GD by of MGD. 

This work is organized as follows. In Sec. \ref{sec2} we present the GD through MGD scheme. In Sec. \ref{sec3} the Buchdahl limit generalities and its extension to anisotropic domains are revisited. Sec. \ref{sec4} is devoted to introduce the extra packing of mass within the arena of MGD and the constraints on the $\alpha$ parameter and the mass--radius ratios are determined for the minimally deformed Tolman IV space--time. Finally, Sec. \ref{sec5} concludes the work.

Throughout the article we shall employ the mostly negative signature $(+,-,-,-)$. 

\section{Anisotropic sources: Gravitational decoupling by MGD}\label{sec2}
In this section we provide a short and comprehensive introduction to GD by 
MDG. For more details see references \cite{r58,r62}, for example. 

Let us consider the Einstein-Hilbert (E-H hereinafter) action describing the gravitational field coupled to a matter field, through minimal coupling-matter principle given by 
\begin{equation}\label{eq1}
S=S_{E-H}+S_{M},    
\end{equation}
where the E-H action reads
\begin{equation}\label{eq2}
S_{E-H}=\frac{1}{2\kappa}\int \sqrt{-g}Rd^{4}x,    
\end{equation}
being $g\equiv det(g_{\mu\nu})$ the determinant of the metric tensor $g_{\mu\nu}$, $R$ the Ricci scalar and $\kappa=8\pi G c^{-4}$. From now on we shall employ relativistic geometrized units where Newton's gravitational constant $G$ and the speed of light are taken to be the unit \i.e, $G=c=1$. The matter sector $S_{M}$ is given by the following general expression
\begin{equation}\label{eq3}
S_{M}=\int \sqrt{-g}\mathcal{L}_{M}d^{4}x,   
\end{equation}
where $\mathcal{L}_{M}$ stands for the Lagrangian-density matter. So, let us write $\mathcal{L}_{M}$ as
\begin{equation}\label{eq4}
\mathcal{L}_{M}=\mathcal{L}_{\bar{M}}+\alpha\mathcal{L}_{X},
\end{equation}
and  consider that the information on  isotropic, anisotropic or charged fluids, among others, are encoded in the Lagrangian-density $\mathcal{L}_{\bar{M}}$ (throughout the text we will use barred quantities to refer the usual material content), while $\mathcal{L}_{X}$ encipher the new matter fields. In principle these incoming fields can be seen as corrections to general relativity \cite{r97}. So, putting all together and taking variations respect to the inverse metric tensor $\delta g^{\mu\nu}$ in (\ref{eq1}), we arrive at the following field equations describing the gravitational--matter interaction
\begin{equation}\label{eq5}
\frac{\delta S}{\delta g^{\mu\nu}}=0 \Rightarrow G_{\mu\nu}\equiv R_{\mu\nu}-\frac{R}{2}g_{\mu\nu}=-8\pi T_{\mu\nu},
\end{equation}
where $G_{\mu\nu}$ is the Einstein's tensor. The general expression for $T_{\mu\nu}$ after variations is given by
\begin{equation}\label{eq6}
T_{\mu\nu}=\underbrace{-2\frac{\delta\mathcal{L}_{\bar{M}}}{\delta g^{\mu\nu}}+g_{\mu\nu}\mathcal{L}_{\bar{M}}}_{\bar{T}_{\mu\nu}}+\alpha\underbrace{\left(-2\frac{\delta\mathcal{L}_{X}}{\delta g^{\mu\nu}}+g_{\mu\nu}\mathcal{L}_{X}\right)}_{\theta_{\mu\nu}}.
\end{equation}
So,
\begin{equation}\label{eq7}
T_{\mu\nu}=\bar{T}_{\mu\nu}+\alpha\theta_{\mu\nu},
\end{equation}
being $\alpha$ a dimensionless parameter. As we are interested in the study of compact structures describing a spherically symmetric and static space-time, we supplement the field equations (\ref{eq5}) with the most general line element, namely
\begin{equation}\label{eq8}
{ds}^{2}={{e}^{\nu}}{{dt}^{2}}-{{e}^{\eta}}{{dr}^{2}}-{r}^{2}({{d\theta}^{2}}+{{sin}^{2}}\theta{{d\phi}^{2}}).   
\end{equation}
The staticity of this space-time (\ref{eq8}) is ensured by considering $\nu$ and $\eta$ as functions of the radial coordinate $r$ only. 
Putting together equations (\ref{eq5}), (\ref{eq7}) and (\ref{eq8}) one obtains the following system of equations
\begin{align}\label{eq9}
e^{-\eta}\left( {\frac {\eta^{{\prime}}}{r}}-\frac{1}{r^2}\right)+\frac{1}{r^2}&=8\pi\left(\bar{T}^{0}_{0}+\alpha\theta^{0}_{0}\right), \\ \label{eq10}
 -e^{-\eta} \left( {\frac {\nu^{{\prime}}}{r}}+\frac{1}{r^2}\right) +\frac{1}{r^2}&=8\pi \left(\bar{T}^{1}_{1}+\alpha\theta^{1}_{1}\right),\\ \label{eq11}
-\frac{ e^{-\eta}}{4} \left(2 \nu^{{\prime\prime}}+\nu^{\prime 2}+2{\frac {\nu^{{\prime}}-\eta^{{\prime}}}{r}}-\nu^{\prime}\eta^{\prime} \right) &=8\pi\left(\bar{T}^{2}_{2}+\alpha\theta^{2}_{2}\right),
\end{align}
where primes denote derivation with respect to the radial coordinate and 
$\bar{T}^{2}_{2}=\bar{T}^{3}_{3}$
in accordance with the spherical symmetry. Moreover, we shall assume that
the source $\bar{T}_{\mu\nu}$ is a perfect fluid which entails $\bar{T}^{1}_{1}=\bar{T}^{2}_{2}$. The corresponding Bianchi's identity (conservation law) $\nabla^{\mu}T_{\mu\nu}=0$ associated with the system (\ref{eq9})-(\ref{eq11}) reads
\begin{eqnarray}\label{eq12}
&&(\bar{T}^{1}_{1})'-\frac{\nu'}{2}\left(\bar{T}^{0}_{0}-\bar{T}^{1}_{1}\right)-\frac{2}{r}+\nonumber\\
&&+
[
(\theta^{1}_{1})'-\frac{\nu'}{2}(\theta^{0}_{0}-\theta^{1}_{1})-\frac{2}{r}(\theta^{2}_{2}-\theta^{1}_{1})
]=0
\end{eqnarray}
Note that Eq. (\ref{eq12}) represents the generalized Tolman-Oppenheimer-Volkoff (TOV) equation driven the hydrostatic equilibrium of the system. 
%In this opportunity we have taken the input matter distribution to be an isotropic fluid configuration given by
%\begin{equation}\label{eq13}
%\bar{T}_{\mu\nu}=\left(\bar{\rho}+\bar{p}\right)u_{\mu}u_{\nu}-\bar{p}g_{\mu\nu}, 
%\end{equation}
%where $\bar{\rho}$ and $\bar{p}$ stand for the energy-density and the pressure, respectively. Moreover, $u^{\beta}=e^{-\nu(r)/2}\delta^{\beta}_{0}$ is the unit time-like four-velocity satisfying $u^{\beta}u_{\beta}=1$. 
At this point one can identify an effective energy density
\begin{equation}
\tilde{\rho}=\bar{T}^{0}_{0}+\alpha\theta^{0}_{0}=\bar{\rho}+\alpha\theta^{0}_{0}, \label{eq14}
\end{equation}
an effective radial pressure
\begin{equation}
\tilde{p}_{r}=-\bar{T}^{1}_{1}-\alpha\theta^{1}_{1}=\bar{p}-\alpha\theta^{1}_{1}\\ ,    \label{eq15}
\end{equation}
and an effective tangential pressure
\begin{equation}
\tilde{p}_{\perp}=-\bar{T}^{2}_{2}-\alpha\theta^{2}_{2}=\bar{p}-\alpha\theta^{2}_{2}.\label{eq16}
\end{equation}

Note that, Eqs. (\ref{eq9}), (\ref{eq10}) and (\ref{eq11}) correspond to a set of differential equations in which we have only separated the components of the matter sector. However, in the framework of MGD these equations can be successfully decoupled and at the end, the system becomes in two set of differential equations, one for each source. To this end, we need to introduce the so--called minimal geometric deformation in the $g^{rr}$
component of the metric given by
the linear map 
\begin{equation}\label{eq17}
e^{-\eta(r)}\mapsto e^{-\eta(r)}=\mu(r)+\alpha f(r).    
\end{equation}

Next, plugging Eq. (\ref{eq17}) into Eqs. (\ref{eq9})--(\ref{eq11}) one arrives to
\begin{eqnarray}\label{ro1}
8\pi\bar{\rho}&=&\frac{1}{r^{2}}-\frac{\mu}{r^{2}}-\frac{\mu^{\prime}}{r}\\\label{p1}
8\pi \bar{p}&=&-\frac{1}{r^{2}}+\mu\left(\frac{1}{r^{2}}+\frac{\nu^{\prime}}{r}\right)\\\label{p2}
8\pi \bar{p}&=&\frac{\mu}{4}\left(2\nu^{\prime\prime}+\nu^{\prime2}+2\frac{\nu^{\prime}}{r}\right)+\frac{\mu^{\prime}}{4}\left(\nu^{\prime}+\frac{2}{r}\right),
\end{eqnarray}
along with the conservation equation
\begin{equation}\label{conservde}
\bar{p}^{\prime}+\frac{\nu^{\prime}}{2}\left(\bar{\rho} +\bar{p}\right)=0,
\end{equation}
for the isotropic sector. Similarly, one has the following equations for the $\theta$-- sector
\begin{eqnarray}\label{cero}
8\pi\theta^{0}_{0}&=&-\frac{f}{r^{2}}-\frac{f^{\prime}}{r} 
\\ \label{one}  
8\pi\theta^{1}_{1}&=&-f\left(\frac{1}{r^{2}}+\frac{\nu^{\prime}}{r}\right)  \\  \label{dos}
8\pi\theta^{2}_{2}&=&-\frac{f}{4}\left(2\nu^{\prime\prime}+\nu^{\prime2}+2\frac{\nu^{\prime}}{r}\right)-\frac{f^{\prime}}{4}\left(\nu^{\prime}+\frac{2}{r}\right). \label{tres}
\end{eqnarray}
The corresponding conservation equation $\nabla^{\nu}\theta_{\mu\nu}=0$ yields
\begin{equation}\label{conservationtheta}
\left(\theta^{1}_{1}\right)^{\prime}-\frac{\nu^{\prime}}{2}\left(\theta^{0}_{0}-\theta^{1}_{1}\right)-\frac{2}{r}\left(\theta^{2}_{2}-\theta^{1}_{1}\right)=0.  
\end{equation}
At this point some comments are in order.
First, note that as can be seen from Eqs. (\ref{eq9})--(\ref{eq11}) the function $\eta(r)$ is quite involved in the full set of equations. So, the only possibility to separate $\{\bar{\rho},\bar{p}\}$ from $\{\theta^{0}_{0},\theta^{1}_{1},\theta^{2}_{2}\}$ is performing the linear map (\ref{eq17}). 
%Then MGD discards deformations on the temporal component of the metric tensor, remaining unchanged. This fact is evident, since equation (\ref{eq9}) shows that the density only depends on the radial component of the metric tensor, which implies that if the deformation is performed on $\nu(r) $ it is not possible then to separate $\bar{\rho}$ from $\theta^{0}_{0}$.\\
Second, as it is well-known the gravitational mass function is defined in terms of $g^{rr}$ as follows
    \begin{equation}\label{mass}
    m(r)=\frac{r}{2}\left[1-e^{-\lambda(r)}\right].    
    \end{equation}
Now, after replacing the linear map (\ref{eq17}) into Eq. (\ref{mass}) the gravitational mass function becomes
    \begin{equation}\label{mass1}
    m(r)=\frac{r}{2}\bigg[1-\mu(r)-\alpha f(r) \bigg],    
    \end{equation}
    from where the total gravitational mass function can be expressed as
    \begin{equation}\label{mass2}
    m(r)=m_{0}(r)-\alpha\frac{r}{2}f(r),  
    \end{equation}
    with $m_{0}(r)$ the mass function of the isotropic system. Hence, it is clear that MGD introduces a modification in the total mass $m_{0}(R)=M_{0}$ of the original system. Even more, if $f(r)>0\ (f(r)<0)$ and $\alpha<0\ (\alpha>0)$ the original total mass $M_{0}$ is increased by a quantity $\alpha Rf(R)/2 $ and as a consequence the mass--radius ratio is also altered, namely
    \begin{equation}\label{u}
    2u(r)=2u_{0}(r)-\alpha f(r).    
    \end{equation}
%Third, Regarding the decoupler function $f(r)$, in order to maintain a regular behaviour of the inner space--time \i.e, free for singularities, $f(r)$ must be vanish at the center of the object, namely $f(0)=0$. 
Finally, it is remarkable that although
%when the linear map (\ref{eq17}) is placed into this complex set of equations, one can collect the seed space--time $\{e^{\nu},\mu\}$ with its corresponding energy--momentum tensor $\{\bar{\rho},\bar{p}\}$ and the new piece $\{\theta^{0}_{0},\theta^{1}_{1},\theta^{2}_{2}\}$ with its associated geometry $f(r)$. The 
the resulting set of equations for the $\theta$--sector (\ref{cero})--(\ref{dos}), correspond to the so--called quasi Einstein field equations in the sense that there is a missing $\frac{1}{r^{2}}$, %. Remarkably, 
Eq. (\ref{conservationtheta}) corresponds to the classical hydrostatic relativistic equilibrium equation associated to $\theta_{\mu\nu}$. \\

An important thing to be noted here, is that the interaction between the two sources is completely gravitational. This fact is reflected by equations (\ref{conservde}) and (\ref{conservationtheta}), where both sectors are individually conserved, namely
%. To be more precise, the Bianchi's identities are satisfied for each tensor after gravitational decoupling, namely 
$\nabla_{\mu}\bar{T}^{\mu\nu}=0$ and $\nabla_{\mu}\theta^{\mu\nu}=0$. 

%On the other hand, in general the deformation (\ref{eq17})
%can be generalized by considering the
%deformation of the $g_{tt}$ component 
%as reported in \cite{r97}. However, this case is out of the %scope of this work.

Note that, when an isotropic solution is considered, Eqs. (\ref{ro1}), (\ref{p1}) and (\ref{p2})
are automatically satisfied. In this respect, we can use the metric function $\nu$ to solve for the $\theta$--sector
which consists in four unknowns, namely $\{\theta^{0}_{0},\theta^{1}_{1},\theta^{2}_{2},f\}$ and only three equations (\ref{cero})-(\ref{dos}). Hence, in order to close the system it is necessary to prescribe some additional information. Regardingly, the so--called mimetic constraints has been broadly used \cite{r62,r63,r64,r65,r66,r67} which 
consist in to impose
\begin{eqnarray}\label{mimic1}
\bar{p}=\theta^{1}_{1},
\end{eqnarray}
or
\begin{eqnarray}\label{mimic2}
\bar{\rho}=\theta^{0}_{0}.
\end{eqnarray}
The former means that the anisotropic behavior enters into the matter distribution through the disturbance of the isotropic pressure of the seed solution. Besides, when (\ref{mimic1}) is imposed from Eqs. (\ref{p1}) and (\ref{one}) one gets
\begin{equation}\label{f1}
f(r)=\frac{1}{1-r\nu^{\prime}(r)}-\mu(r).    
\end{equation}
%As was discussed earlier, the decoupler function $f(r)$ must fulfill some requirements in order avoid some pathologies inside the stellar interior. So, the decoupler function $f(r)$ must be null at $r=0$ in order to maintain $e^{-\lambda(0)}=1$ as it is required for all well behaved stellar interior. 
It is worth noticing that restriction (\ref{mimic1}) entails an interesting situation in considering the total mass contained by the fluid sphere. As said before, MGD modifies the gravitational mass function (\ref{mass2}), and as a consequence the total mass $m(R)=M$ contained by the object. Surprisingly, when (\ref{mimic1}) is used, the extra piece $\alpha rf(r)/2$ vanishes at the boundary $r=R$, which allows to conclude that this mimic constraint does not modify the total mass. Indeed, a close look reveals that Eq. (\ref{p1}) at $r=R$ provides 
\begin{equation}
R\nu^{\prime}(R)=\frac{1}{\mu(R)}-1,
\end{equation}
which after being replaced into (\ref{f1}) evaluated at the surface, it is obtained $f(R)=0$. From the physical point of view this result can be interpreted as a redistribution of the mass of the object.\\
The situation is quite different when the mimetic constraint for the density (\ref{mimic2}) is employed.
In this case, after replacing
(\ref{ro1}) and (\ref{one}) we obtain
\begin{eqnarray}\label{f2}
f(r)=\mu(r)-1.
\end{eqnarray}
Now, as we are dealing with a spherical mass distribution the total mass is computed as follows
\begin{equation}
m(R)=4\pi\int^{R}_{0}r^{2}\rho dr=4\pi\int^{R}_{0}r^{2}\left(\bar{\rho}+\alpha\theta^{0}_{0}\right) dr,   
\end{equation}
where by virtue of (\ref{mimic2}) the above expression becomes
\begin{equation}
m(R)=4\pi\int^{R}_{0}r^{2}\left(1+\alpha\right)\bar{\rho} dr=\left(1+\alpha\right)m_{0}(R),
\end{equation}
from where it is evident that if $\alpha>0$ then $m(R)>m_{0}(R)$. 
%In this case the anisotropic behaviour rises into the system by altering the isotropic density $\bar{\rho}$. Moreover, by equating (\ref{ro1}) and (\ref{cero}) the deformation function $f(r)$ has the following form
%\begin{equation}\label{f2}
%f(r)=\mu(r)-1.    
%\end{equation}
As we shall see later,
in this work we base our analysis of extra packing of mass in the Tolman IV solution extended by MGD 
using the mimic constraint for the density previously reported in \cite{r62}.
%as we shall see in what follows. 
%It is worth mentioning that this methodology has been used also to extend the Tolman IV solution.

\section{ Buchdahl's limit: Isotropic and anisotropic sources revisiting}\label{sec3}

Buchdahl's limit states that for a spherically symmetric and static configuration describing an isotropic matter distribution, the maximum mass--radius ratio $u$ (also known as compactness factor) is given by
\begin{equation}\label{Blimit}
u\equiv \frac{M}{R}\leq \frac{4}{9},
\end{equation}
where $M$ and $R$ stand for the gravitational mass contained in the sphere and the radius of the star, respectively. As can be seen from (\ref{Blimit}) one has two options i) an equality and ii) an inequality. The first option holds by considering a constant energy density (incompressible fluid), a degenerate metric (
in the sense that $g_{tt}(r=0)=0$) and a non well behaved pressure (divergent pressure at the center of the compact configuration). 
Otherwise, the inequality requires a positive and
monotonously decreasing energy density from the center to the boundary of the compact object
density everywhere inside the star and a vanishing pressure at the surface of the structure. Hence, the value $4/9$ is  an absolute upper limit for all static fluid spheres whose density does not increase outwards. The condition (\ref{Blimit}) works in the isotropic case but as it was said before celestial bodies are not necessarily made up by perfect fluid distributions.
%Relaxing the isotropic condition and allowing the presence of local anisotropies in the stellar interior new constraints arise on the compactness factor. Moreover, some modifications on relevant physical quantities such as the surface gravitational redshift $z_{s}$ are induced. 
In this respect B\"{o}hmer and Harko \cite{r43} derived the corresponding upper bound for the mass--radius relation for an anisotropic matter distribution in presence of a cosmological constant $\Lambda$ obtaining 
\begin{equation}\label{upper}
\frac{2M}{R}\leq \left(1-\frac{8\pi}{3}\lambda R^{2}\right) \left[1-\frac{1}{9}\frac{\left(1-2\lambda/\langle\rho\rangle\right)^{2}}{\left(1-\frac{8\pi}{3}\lambda R^{2}\right)\left(1+F\right)^{2}}\right].    
\end{equation}
In the above expression $\langle\rho\rangle$ stands for the mean energy density and $F$ is a
function proportional to the anisotropy factor $\Delta=p_{\bot}-p_{r}$, which is given by
\begin{equation}
F=2\frac{\Delta(R)}{\langle\rho\rangle}\Bigg[\frac{\arcsin\left(\sqrt{\frac{2M\chi\left(R\right)}{R}}\right)}{\sqrt{\frac{2M\chi\left(R\right)}{R}}}-1\Bigg],
\end{equation}
where $\chi$ is given by
\begin{equation}
\chi(r)\equiv 1+\frac{4\pi}{3}\Lambda\frac{r^{3}}{m(r)}. 
\end{equation}
As we are interested in studying space-time without cosmological constant from now on we will set off $\Lambda$. Then the upper bound (\ref{upper}) becomes 
\begin{eqnarray}\label{mbl}
2u\leq 1-\frac{1}{9 (1+F)^2},
\end{eqnarray}
and $F$ turns
\begin{eqnarray}\label{F}
F=\frac{R^2 \Delta (u)}{\frac{3 u}{8 \pi }}\left(\frac{\sin ^{-1}\left(\sqrt{2 u}\right)}{\sqrt{2 u}}-1\right).
\end{eqnarray}
 At this point a couple of comments are in order. First, 
note that $F$ is a positive quantity. Otherwise, 
the condition $\Delta\ge0$ could be violated. As a consequence, non--vanishing values of $F$, Eq. (\ref{mbl}) allows extra packing of mass in compact stellar structures. Second, from Eq. (\ref{F}), it is straightforward to show that the bounds of the compactness parameter of the anisotropic distribution, $u$, is defined in the interval 
\begin{eqnarray}\label{ub}
\frac{4}{9}\le u< \frac{1}{2}.
\end{eqnarray}
In the next section we will discuss in details how the local anisotropies introduced in the stellar interior by gravitational decoupling through MGD approach contributes on the maximum  mass-radius ratio value allowable for relativistic anisotropic fluid spheres in the arena of GR.

\section{Local anisotropy induced by MGD}\label{sec4}
In the context of MGD, the anisotropy of the total matter sector seeded by a perfect fluid, can be written as
\begin{eqnarray}\label{delalpha}
\Delta=p_{\perp}-p_{r}=\alpha(-\theta^{2}_{2}-(-\theta^{1}_{1})). 
\end{eqnarray}
%Note that, in order to satisfy the requirement $\Delta>0$, we have to impose extra constraints, for example, $-\theta^{2}_{2}-(-\theta^{1}_{1})>0$ and $\alpha\ge0$. Now, as commented at the end of the previous section, the introduction of local anisotropies could lead to extra packing of mass in the interior of a compact object. In this sense, given the link
%between the anisotropy and the decoupler matter content (see Eq. (\ref{delalpha})), the gravitational decoupling of sources by MGD can be thought as a
%kind of mechanism that allows to introduce such an extra packing.
%Another aspect that deserves to be pointed out, 
Now, from Eq. (\ref{mbl}) it is straightforward to see that
%is that
the connection between the anisotropy induced by gravitational decoupling and the compactness parameter is given by
\begin{equation}\label{blmgd}
2 u\leq 1-\frac{1}{9 \left(1+\alpha\frac{16 \pi    R^2 (\theta^{1}_{1}-\theta^{2}_{2})}{3 u}\left(\frac{\sin ^{-1}\sqrt{2 u}}{\sqrt{2 u}}-1\right)\right)^2},
\end{equation}
which leads to the classic
Buchdahl's limit for the isotropic case when $\alpha\to0$, as expected.  However, expression (\ref{blmgd}) must be considered as a formal expression because matching conditions namely, the continuity of the first and the second fundamental form of the total solution leads to non--linear equations involving $\alpha$ and $u$. In this sense, obtaining an analytical expression of the bound of the compactness parameter of the total solution, $u$, in terms of the decoupling parameter is not possible. Instead, we can try to find the connection between $u$ and $\alpha$ in an alternative manner. To this end, we shall consider the MGD--extended Tolman IV solution previously reported in  \cite{r62}. As it is well known, the Tolman IV solution that solve the isotropic sector parametrized by $(\nu,\mu,\rho,p)$ is given by
\begin{eqnarray}\label{mu}
e^{\nu}&=&B^{2}(1+\frac{r^{2}}{A^{2}}),\\ \label{nui}
\mu&=&\frac{\left(1-\frac{r^{2}}{C^{2}}\right)\left(1+\frac{r^{2}}{A^{2}}\right)}{1+\frac{2r^{2}}{A^{2}}},\\
\rho&=&\frac{3A^{4}+A^{2}(3C^{2}+7r^{2})+2r^{2}(C^{2}+3r^{2})}{8\pi C^{2}(A^{2}+2r^{2})^{2}},\\
p&=&\frac{C^{2}-A^{2}-3r^{2}}{8\pi C^{2}(A^{2}+2r^{2})},
\end{eqnarray}
where $A,B$ and $C$ are constants. Now, using $\nu$ given by Eq. (\ref{mu}) the extended solution can be obtained by imposing extra conditions to close the system (\ref{cero})--(\ref{tres}).
%of differential equations.\\
As it was  previously commented, in this work we shall concentrate in the so--called mimic constraint for the density which after
replacing (\ref{nui}) in (\ref{f2})
leads to
\begin{equation}
f(r)=-\frac{r^{2}}{C^{2}}
\left(
\frac{A^{2}+C^{2}+r^{2}}{A^{2}+2r^{2}}
\right).\label{ff}   
\end{equation}
Note that, at this point the $\theta$--sector can be completely determined in terms of the decoupling function given by (\ref{ff}). In particular, a strightforward computations reveals that
the anisotropy function of the extended solution can be written as \cite{r62}
\begin{eqnarray}
\Delta(r;\alpha)=\alpha\left(\theta^{1}_{1}-\theta^{2}_{2}\right)=\frac{\alpha r^{2}}{8\pi(A^{2}+r^{2})^{2}}.
\end{eqnarray}
Now, with all of this results at hand we can obtain the source of the extended solution, namely 
\begin{eqnarray}
\tilde{\rho}&=&(\alpha +1) \bar{\rho},\\
\tilde{p}_{r}&=&
\bar{p}-\frac{\alpha  \left(\left(A^2+3 r^2\right) \left(A^2+C^2+r^2\right)\right)}{8 \pi  C^2 \left(A^2+r^2\right) \left(A^2+2 r^2\right)},\\
\tilde{p}_{\perp}&=&
\tilde{p}_{r}+
\frac{\alpha  r^2}{8 \pi  \left(A^2+r^2\right)^2}.
\end{eqnarray}

In order to proceed with the study of the modification of the Buchdahl limit induced by MGD we need to compare the compactness parameter of the original isotropic solution, $u_{0}=\frac{M_{0}}{R}$, and
the compactness parameter
of the anisotropic solution given by $u=\frac{M}{R}$. Incidentally, in Ref. \cite{r62} it was shown that
this quantities are related by
\begin{eqnarray}\label{rel}
2u=2u_{0}+\alpha\frac{R^{2}}{C^{2}}
\left(
\frac{A^{2}+C^{2}+R^{2}}{A^{2}+2R^{2}}
\right).
\end{eqnarray}
At this point a couple of comments are in order. First, note that all the quantities involved in the previous expression are positive which entails that $u>u_{0}$. Second, it is noticeable that
when the Buchdahl limit is saturated in the isotropic sector, namely
$u_{0}= \frac{4}{9}$, the anisotropic solution acquire an extra packing in the sense that
\begin{eqnarray}
u\ge \frac{4}{9}.
\end{eqnarray}
However, note that the above corresponds to the critical situation in which $u_{0}$ acquires its maximum allowed value. In this sense, we may wonder how the parameters should be set to ensure extra packing of mass in more realistic situations where $u_{0}< 4/9$. 
It is worth mentioning that such a set is not trivial as we shall see in what follows.

%So, to explore some realistic limits of this toy model and assures an extra packing of mass, it is necessary 
The strategy we shall employ here corresponds to express the set of arbitrary parameters $\{A,B,C\}$ in terms of $\{R,u,u_{0},\alpha\}$ an then
we demand
\begin{eqnarray}
A>0\label{ca}\\
B>0\label{cb}\\
C>0\label{cc}.
\end{eqnarray}
for some set $\{u,u_{0},\alpha,R\}$ such that
\begin{eqnarray}
u_{0}< \frac{4}{9}\label{c1}\\
\frac{4}{9}<u<\frac{1}{2}\label{c2}\\
\alpha>0\label{c3}\\
R>0.
\end{eqnarray}
In order to do so, we shall employ the Israel--Darmois junction conditions \cite{is,dar} which consist in to demand 
%.This mechanism invokes
the continuity of the first and second fundamental forms
across the boundary $\Sigma:r=R$ of the star. Now, for the continuity  of 
the first fundamental form %consists in the continuity of the metric potentials across the boundary $\Sigma:r=R$ of the star. To accomplish this, 
it is required that the inner manifold $\mathcal{M}^{-}$ is joined in a smoothly way with the outer space--time $\mathcal{M}^{+}$ described by the vacuum Schwarzschild solution
\begin{equation}
ds^{2}=\left(1-2\frac{M_{\text{Sch}}}{r}\right)dt^{2}-\left(1-2\frac{M_{\text{Sch}}}{r}\right)^{-1}dr^{2}-r^{2}d\Omega^{2}.   
\end{equation}
Hence,
\begin{equation}
g^{-}_{tt}\bigg{|}_{r=R}=g^{+}_{tt}\bigg{|}_{r=R} \quad \mbox{and} \quad g^{-}_{rr}\bigg{|}_{r=R}=g^{+}_{rr}\bigg{|}_{r=R}.    
\end{equation}
For the present situation we have
\begin{eqnarray}\label{fft}
B^{2}(1+\frac{R^{2}}{A^{2}})&=&1-2\frac{M}{R},\\ \label{ffr}
\left(1+\alpha\right)\frac{\left(1-\frac{R^{2}}{C^{2}}\right)\left(1+\frac{R^{2}}{A^{2}}\right)}{1+\frac{2R^{2}}{A^{2}}}-\alpha&=&1-2\frac{M}{R},
\end{eqnarray}
where at $\Sigma$ the Schwarzschild mass $M_{\text{Sch}}$ coincides with the total mass $M$ contained by the fluid sphere. The second fundamental form entails a vanishing radial pressure $\tilde{p}_{r}$ at the surface of the compact structure. Explicitly it reads
\begin{equation}\label{sf}
\tilde{p}_{r}(R)=0.    
\end{equation}
Now, from Eqs. (\ref{rel}) and (\ref{fft}) we get
\begin{eqnarray}
    B&=&\sqrt{
\frac{1-2 u_{0}-\alpha\frac{R^2}{C^2}\left(\frac{A^2+C^2+R^2}{A^2+2R^2}\right)}{1+\frac{R^2}{A^2}}
   } \label{B},
\end{eqnarray}
and from (\ref{sf})
\begin{equation}
    C=\sqrt{\frac{(1+\alpha ) \left(A^2+R^2\right) \left(A^2+3
   R^2\right)}{\left(A^2+R^2\right)-\alpha  \left(A^2+3 R^2\right)}}. \label{C}
\end{equation}
Next, using (\ref{C}) in  (\ref{rel}) we obtain
\begin{equation}
\label{A}
    A=\frac{\sqrt{\alpha-3u(1+\alpha)+3u_{0}(1+\alpha)}}{\sqrt{(u-u_{0})(1+\alpha)}}.
\end{equation}
%It should be noted that Eqs. (\ref{rel}) and (\ref{ffr}) are equivalents.
 
Without loss of generality we shall set $R=1$ \footnote{It is clear the the same analysis can be performed for arbitrary $R$. However, the expressions are not illuminating at all.} in what follows, so Eqs. (\ref{B}) and (\ref{C})  read
\begin{align}
B&=\sqrt{\frac{(2 u-1) (-\alpha +3 (\alpha +1) u-3 (\alpha +1) u_{0})}{\alpha -2 (\alpha +1) u+2 (\alpha +1) u_{0}}}\\
C&=\sqrt{\frac{\alpha  (-\alpha +2 (\alpha +1) u-2 (\alpha +1) u_{0})}{(u-u_{0}) ((\alpha -1) \alpha +2 (\alpha +1) u-2 (\alpha +1) u_{0})}}.
\end{align}
Now, Eqs. (\ref{ca}), (\ref{cb}), (\ref{cc}), constrained by (\ref{c1}), (\ref{c2}) and (\ref{c3}) reduce to following extra conditions
\begin{eqnarray}
3 (\alpha +1) u<\alpha +3 (\alpha +1) u_{0}\label{cc1}\\
2 (\alpha +1) u<\alpha +2 (\alpha +1) u_{0}\label{cc2}\\
\alpha ^2+2 (\alpha +1) u<\alpha +2 (\alpha +1) u_{0}\label{cc3},
\end{eqnarray}
from where $u_{0}$ and $\alpha$ acquire extra constraints, namely
\begin{eqnarray}
u_{0}&>&0.38889\label{u0e}\\
0.2<\alpha&<&0.333\label{alfae}
\end{eqnarray}
At this point some comments are in order. First, it is worth mentioning that (\ref{u0e}) states a lower bound for the compactness parameter of the perfect fluid solution. Even more, for values out the interval
\begin{eqnarray}
0.38889<u_{0}<\frac{4}{9},
\end{eqnarray}
the anisotropic compactness $u$ can not surpass the Buchdahl limit for isotropic solutions and
and the extra packing of mass is not possible in this case. Second, the extra packing condition allows to restrict the appropriate values for the decoupling parameter as shown in (\ref{alfae}). Finally, it is worth mentioning that constraints given by Eqs. 
(\ref{u0e}) and (\ref{alfae}) correspond to a particular case we obtained when the system
(\ref{cc1}), (\ref{cc2}) and (\ref{cc3}) is reduced. However, we considered the the most simplest case in order to illustrate 
how the  process works.

\begin{figure}[h!]
    \centering
\includegraphics[scale=0.45]{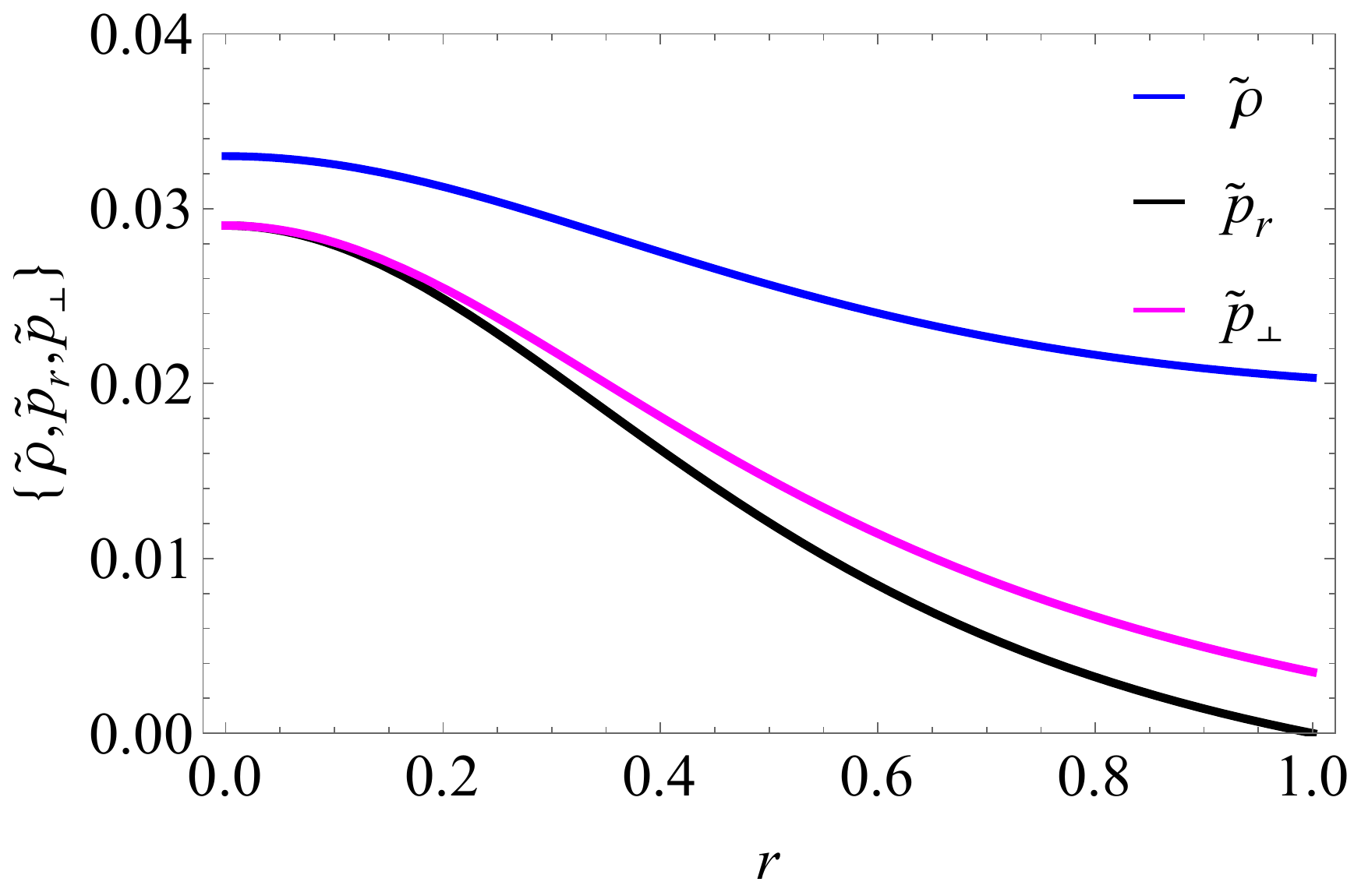}
\caption{Decreasing behavior of density and pressures of the anisotropic solution.}
\label{rhopr}
\end{figure}

As an example we shall study the behaviour of a solution with extra packing of mass considering $u=0.45$, $u_{0}=0.39$ and $\alpha=0.3$. In figure \ref{rhopr} it is shown the behavior of the density, and the effective radial $\tilde{p}_{r}$ and tangential $\tilde{p}_{\perp}$ pressures of the anisotropic solution. Note that all the quantities shown in the profiles are monotonously decreasing and reach their maximum value at the center of the star, as expected. Besides, the radial $\tilde{p}_{r}$ pressure vanishes at the surface of the star.

Now we analyse the stability of the solution using the adiabatic index criteria. 
As it is well known, the radial and the tangential adiabatic index, $\Gamma_{r}$ and $\Gamma_{\perp}$ respectively, defined as 
\begin{eqnarray}
\Gamma_{r}&=&\frac{\tilde{\rho}+\tilde{p}_{r}}{\tilde{p}_{r}}\frac{d\tilde{p}_{r}}{d\tilde{\rho}}\\
\Gamma_{\perp}&=&\frac{\tilde{\rho}+\tilde{p}_{\perp}}{\tilde{p}_{\perp}}\frac{d\tilde{p}_{\perp}}{d\tilde{\rho}},
\end{eqnarray}
allow to connect the relativistic structure of a spherical static object and the equation of state of the interior fluid and serves as a criterion of stability. More precisely, it is said that an interior configuration is stable whenever $\Gamma_{r}$ and $\Gamma_{\perp}$ are grater that $4/3$ for static equilibrium. Otherwise, it is said that the object is unstable and the self gravitating object undergoes a gravitational collapse.

\begin{figure}[h!]
    \centering
\includegraphics[scale=0.45]{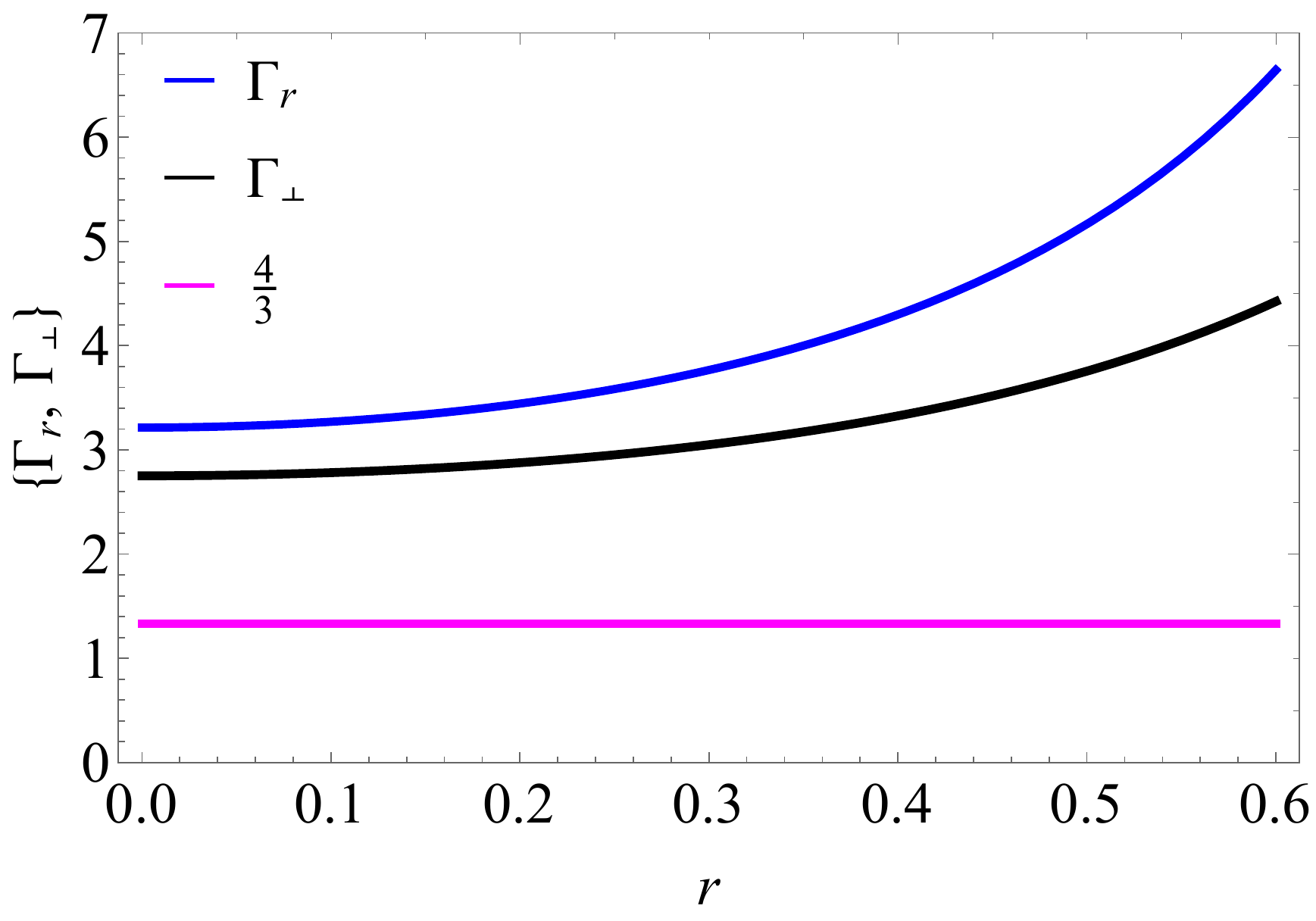}
\caption{Adiabatic index showing stability for the anisotropic interior solution case (greater than $4/3$).}
\label{adiabatic}
\end{figure}

As shown in figure \ref{adiabatic}, the behaviour of the adiabatic index reveals that for $u=0.45$, we obtain a stable interior configuration from the anisotropic solution. Furthermore, note that $\Gamma_{r}>\Gamma_{\perp}$ in accordance with the monotonously increasing of the anisotropy given that given that $p_{r}\ge p_{\perp}$.
In this sense, we conclude that MGD method not only allow to extend isotropic solutions to anisotropic domains but it can be used to obtain a physical acceptable interior with extra packing of mass. Furthermore, the extra packing condition leads to
modifications in the interval of the compactness parameter of the isotropic solution. More precisely, given that $0.38889<u_{0}<\frac{4}{9}$, not any arbitrary isotropic solution allows being extended with extra packing but only those with a high compactness comparable with observational data of neutron stars. Finally, it should be emphasized that the extra packing condition leads also to a stringent restriction in the decoupling parameter $\alpha$ as shown in Eq. (\ref{alfae}).

%%%%%%%%%%%%%%%%%%%%%%%%%%%%%%%%%%%%%%%%%%%%%%%%%%%%%%%%%%%%%%%%%%%%%%

\section{Concluding Remarks}\label{sec5}
In this work we implemented the
Gravitational Decoupling in the framework of the
Minimal Geometric Deformation approach to induce extra packing of mass in stellar interiors. As anisotropic solution we considered the Tolman IV extended by using the mimetic constraint for the density. We obtained that although the induced anisotropy enters naturally in the B\"{o}mer and Harko criteria to induce extra packing, obtaining an analytical relationship between the decoupling parameters and the compactness of the star is not possible. However, we were able to set the space of parameters in an appropriate manner to obtain a well posed anisotropic solution with extra packing. We conclude that the Minimal Geometric Deformation approach allows to extend isotropic solutions to anisotropic interiors with extra packing of mass. Besides, the extra packing condition induces a lower bound on the compactness parameter of the isotropic system used as a seed and the decoupling parameter gets a restriction within a well defined interval.

%%%%%%%%%%%%%%%%%%%5
%%%%%%%%%%%%%%%%%%%%%55
\section*{ACKNOWLEDGEMENTS}
F. Tello-Ortiz thanks the financial support by the CONICYT PFCHA/DOCTORADO-NACIONAL/2019-21190856 and projects ANT-1856 and SEM 18-02 at the Universidad de Antofagasta, Chile, and grant Fondecyt No. $1161192$, Chile and to TRC project-BFP/RGP/CBS/19/099 of the Sultanate of Oman.

%%%%%%%%%%%%%%%%%%%%%%%%%%%%%%%%%%%%%%%%%%%%%%%%%%%%%%%%%%%%%%%%%%%%%%%


\begin{thebibliography}{99}

\bibitem{r1} R. C. Tolman,
Phys. Rev. \textbf{55}, 364 (1939).

\bibitem{r2} H. A. Buchdahl, Phys. Rev. D \textbf{116}, 1027 (1959).

\bibitem{r3} R. Ruderman, Ann. Rev. Astron. Astrophys. \textbf{10}, 427 (1972).

%%%%%%%%%%%%%%%
\bibitem{muk1} U. Das and B. Mukhopadhyay, Phys. Rev. D {\bf 86}, 042001
(2012).
\bibitem{muk2} U. Das and B. Mukhopadhyay, Phys. Rev. Lett. {\bf 110},
071102 (2012).

\bibitem{muk3} U. Das and B. Mukhopadhyay, Int. J. Mod. Phys. D {\bf 21},
1242001 (2012).

\bibitem{muk4} A. Kundu B. Mukhopadhyay, Mod. Phys. Lett. A {\bf 27},
1250084 (2012).

\bibitem{muk5} U. Das, B. Mukhopadhyay and A. R. Rao, Astrophys. J. {\bf 767}, L14 (2013).

\bibitem{muk6} U. Das and B. Mukhopadhyay, Int. J. Mod. Phys. D {\bf 22}, 1342004 (2013).

\bibitem{muk7} U. Das and B. Mukhopadhyay, Mod. Phys. Lett. A {\bf 29},
1450035 (2014).

%%%%%%%%%%%%%%
\bibitem{r4} R. L. Bowers, E. P. T. Liang, Astrophys. J. \textbf{188}, 657 (1974).

\bibitem{r5} J. R. Oppenheimer and G. M. Volkoff,  Phys. Rev. {\bf 55}, 374 (1939).

%%%%%%%%%%%%%%%%%%%%%%%%%%%%%%%%%%%%%%

\bibitem{r6} M. K. Mak, P. N. Dobson, Jr., and T. Harko,  Mod. Phys. Lett. A {\bf15}, 2153 (2000).

\bibitem{r7} H. Andreasson, C. G. Boehmer and A. Mussa, Class. Quant. Grav. {\bf29}, 095012 (2012).

\bibitem{r8} Z. Stuchlik, Acta Phys. Slov. {\bf 50}, 219 (2000).

\bibitem{r9} J. Ponce de Leon and N. Cruz,  Gen. Rel. Grav. {\bf32}, 1207 (2000).

\bibitem{r10} C. Germani and R. Maartens,  Phys. Rev. D {\bf64}, 124010 (2001). 

\bibitem{r11} M. A. Garca-Aspeitia and L. A. Urea-López, Class.
Quant. Grav. {\bf32}, 025014 (2015).

\bibitem{r12} T. Harko and M. K. Mak,  J. Math.
Phys. {\bf41}, 4752 (2000).

\bibitem{r13} T. Singh and L. N. Rai, Gen. Rel. Grav. {\bf15}, 875 (1983).

\bibitem{r14} K. Yokoi, Prog. Theor. Phys. {\bf48}, 1760 (1972).

\bibitem{r15} P. Burikham, T. Harko, and M. J. Lake, Phys. Rev. D {\bf94},  064070 (2016).

\bibitem{r16} S.-e. Tsuneishi, K. Watanabe, and T. Tsuchida,  Prog. Theor. Phys. {\bf115}, 487{522 (2006).

\bibitem{r17} M. W. Horbatsch and C. P. Burgess, JCAP {\bf1108}, 027 (2011). 

\bibitem{r18} P. Pani, E. Berti, V. Cardoso, and J. Read, Phys. Rev. D {\bf84}, 104035 (2011).

\bibitem{r19} T. Tsuchida, G. Kawamura, and K. Watanabe, Prog. Theor. Phys. {\bf100}, 291 (1998).

\bibitem{r20} R. Goswami, S. D. Maharaj, and A. M. Nzioki, Phys. Rev. D {\bf92},  064002 (2015).

\bibitem{r21} U. Das and B. Mukhopadhyay, JCAP \{bf1505}, 045 (2015).

\bibitem{r22} M. Wright, Gen. Rel. Grav. {\bf48}, 93 (2016).

\bibitem{r23} N. Dadhich, A. Molina, and A. Khugaev, Phys. Rev. D {\bf81}, 104026 (2010).

\bibitem{r24} N. Dadhich and S. Chakraborty, Phys. Rev. D {\bf95}, 064059 (2017).

\bibitem{r25} S. H. Hendi, G. H. Bordbar, B. Eslam Panah and S. Panahiyan, JCAP {\bf1707}, 004 (2017).

\bibitem{r26} N. Dadhich, S. Hansraj and B. Chilambwe, Int. J.
Mod. Phys. D {\bf26}, 1750056 (2016).

\bibitem{r27} A. Molina, N. Dadhich and A. Khugaev, Gen. Rel. Grav. {\bf49}, 96 (2017).

\bibitem{r28} S. Nojiri and S. D. Odintsov,  Phys. Rept. {\bf505}, 59 (2011).

\bibitem{r29} T. P. Sotiriou and V. Faraoni, Rev. Mod. Phys. {\bf82}, 451 (2010).

\bibitem{r30} S. Nojiri and S. D. Odintsov, Phys.Rev. D {\bf68}, 123512 (2003).

\bibitem{r31} A. De Felice and S. Tsujikawa, Living Rev. Rel. {\bf13}, 3 (2010).

\bibitem{r32} S. Chakraborty and S. SenGupta, Eur. Phys. J. C {\bf76}, 552 (2016).

\bibitem{r33} T. Padmanabhan and D. Kothawala, Phys. Rept. {\bf531},
115 (2013).

\bibitem{r34} S. Chakraborty, JHEP {\bf08}, 029 (2015).

\bibitem{r35} N. Dadhich, Pramana {\bf74}, 875 (2010).

\bibitem{r36} D. Kastor, Class. Quant. Grav. {\bf29}, 155007 (2012).

\bibitem{r37} C. Breu and L. Rezzolla, 
Mon. Not. Roy. Astron. Soc. {\bf459}, 646 (2016).

\bibitem{r38} A. Fujisawa, H. Saida, C.-M. Yoo and Y. Nambu, Class. Quant. Grav. {\bf32}, 215028 (2015).

\bibitem{r39} D. Barraco and V. H. Hamity, Phys.
Rev. D {\bf65}, 124028 (2002).

\bibitem{r40} P. Karageorgis and J. G. Stalker, Class. Quant.
Grav. {\bf25}, 195021 (2008).

\bibitem{r41} H. Andreasson, J. Diff. Eq.
{\bf245}, 2243 (2008).

\bibitem{r42} Sumanta Chakraborty and Soumitra SenGupta, JCAP {\bf05}, 032 (2018).

\bibitem{r43} C. G. B\"{o}hmer and T. Harko, Class. Quant. Grav. {\bf23}, 6479 (2006).

%\bibitem{r44} L. Herrera, N. O. Santos, 
%Phys. Rep. \textbf{286}, 53 (1997).

%\bibitem{r45} M. K. Mak, T. Harko,  Proc.Roy.Soc.Lond. A \textbf{459}, 393 (2003).

\bibitem{r58} J. Ovalle. Phys. Rev. D {\bf 95}, 104019 (2017). 

\bibitem{r62} J. Ovalle,  R. Casadio, R. da Rocha, A. Sotomayor. Eur. Phys. J. C {\bf 78}, 122 (2018).

\bibitem{r97} J. Ovalle, Phys. Lett. B {\bf 788}, 213 (2019). 


\bibitem{r46} J. Ovalle. Mod. Phys. Lett. A {\bf 23}, 3247 (2008).

\bibitem{new1} J. Ovalle, \emph{Gravitation and Astrophysics} (ICGA9), Ed. J. Luo, World Scientific, Singapore, 173-182 (2010).

\bibitem{r53} R. Casadio, J. Ovalle, and R. da Rocha, \emph{Class. Quantum Grav.} {\bf 32}, 215020 (2015).

\bibitem{r56} J. Ovalle. Int. J. Mod. Phys. Conf. Ser. {\bf 41}, 1660132 (2016).

\bibitem{r47} J. Ovalle. Int. J. Mod. Phys. D {\bf 18}, 837 (2009).

\bibitem{r48} J. Ovalle. Mod. Phys. Lett. A {\bf 25}, 3323 (2010).

\bibitem{r49} R. Casadio, J. Ovalle. Phys. Lett. B {\bf 715}, 251 (2012).

\bibitem{r50} J. Ovalle and F. Linares \emph{Phys. Rev. D} {\bf 88}, 104026 (2013).

\bibitem{r51} J. Ovalle, F. Linares, A. Pasqua and A. Sotomayor. \emph{Class. Quantum Grav.} {\bf 30}, 175019 (2013).

\bibitem{r52} R. Casadio, J Ovalle and R. da Rocha, \emph{Class. Quantum Grav.} {\bf 30}, 175019 (2014).

\bibitem{r54} J. Ovalle, L. A. Gergely and R. Casadio, \emph{Class. Quantum Grav.} {\bf 32}, 045015 (2015).

\bibitem{r55} R. Casadio, J. Ovalle and R. da Rocha, \emph{Europhys. Lett.} {\bf 110}, 40003 (2015).

\bibitem{r57} R. Casadio and R. da Rocha, \emph{Phys. Lett. B} {\bf 763}, 434 (2016).

\bibitem{r59} R. da Rocha, Phys. Rev. D {\bf 95}, 124017 (2017).

\bibitem{r60} R. da Rocha, \emph{Eur. Phys. J. C} {\bf 77}, 355 (2017).

\bibitem{r61} R. Casadio, P. Nicolini and R. da Rocha, \emph{Class. Quantum Grav.} {\bf 35}, 185001 (2018). 

\bibitem{r63} M. Estrada, F. Tello-Ortiz. Eur. Phys. J. Plus {\bf 133},  453 (2018) .

\bibitem{r64} C. Las Heras, P. Leon. Fortschr. Phys. {\bf 66}, 1800036 (2018).

\bibitem{r65} L. Gabbanelli,  A. Rinc\'on, C. Rubio. Eur. Phys. J. C {\bf 78}  370 (2018).

\bibitem{r66} M. Sharif, S. Sadiq,  Eur. Phys. J. C {\bf 78}, 410 (2018).

\bibitem{r67} E. Morales, F. Tello-Ortiz, Eur. Phys. J. C {\bf 78}, 618 (2018).

\bibitem{r68} S. Maurya, F. Tello, Eur. Phys. J. C {\bf 79}, 85 (2019)

\bibitem{r69} E. Morales, F. Tello-Ortiz,  Eur. Phys. J. C{\bf 78}, 841 (2018).

\bibitem{r70} G. Abell\'an, A. Rinc\'on, E. Fuenmayor and E. Contreras, arXiv:2001.07961 

\bibitem{r71} G. Abell\'an, V. Torres, E. Fuenmayor and E. Contreras, Eur. Phys. J. C {\bf 80}, 177 (2020). 

\bibitem{r72} J. Ovalle, R. Casadio, R. da Rocha, A. Sotomayor, Z. Stuchlik, Eur. Phys. J. C {\bf 78}, 960 (2018).

\bibitem{r73}J. Ovalle,  R. Casadio, R. da Rocha, A. Sotomayor and Z. Stuchlik, EPL {\bf 124}, 20004 (2018).

\bibitem{r74} M. Sharif, S. Saba, Eur. Phys. J. C {\bf 78}, 921(2018).

\bibitem{r75} M. Sharif, S. Sadiq,  Eur. Phys. J. Plus {\bf 133}, 245 (2018).

\bibitem{r76} A. Fernandes-Silva, A. J. Ferreira-Martins, R. da Rocha. Eur. Phys. J. C {\bf 78}, 631 (2018).

\bibitem{r77}A. Fernandes-Silva, R. da Rocha. Eur. Phys.J. C {\bf 78}, 271 (2018).

\bibitem{r78} E. Contreras and P. Bargue\~no. Eur. Phys. J. C {\bf 78}, 558 (2018). 

\bibitem{r79} G. Panotopoulos, \'A. Rinc\'on, Eur. Phys. J. C {\bf 78}, 851 (2018)

\bibitem{r80} E. Contreras and P. Bargue\~no, Eur. Phys. J. C {\bf 78}, 985 (2018).

\bibitem{r81}M. Estrada, R. Prado, Eur. Phys. J. Plus {\bf 134}, 168 (2019). 

\bibitem{r82} E. Contreras, Class. Quantum Grav {\bf 36}, 095004 (2019).

\bibitem{r83}  E. Contreras, \'A. Rinc\'on and P. Bargue\~no, Eur. Phys. J. C {\bf 79}, 216 (2019) .

\bibitem{r84}E. Contreras and P. Bargue\~no, Class. Quantum Grav. 36, 215009 (2019).

\bibitem{r85} S. Maurya, and F. Tello-Ortiz, Phys. Dark Univ. {\bf27}, 100442 (2020).

\bibitem{r86} M. Estrada,  Eur. Phys. J. C {\bf79}, 918 (2019).

\bibitem{r87} L. Gabbanelli, J. Ovalle, A. Sotomayor, Z. Stuchlik, R. Casadio, Eur. Phys. J. C {\bf 79}, 486 (2019).

\bibitem{r88} J. Ovalle, C. Posada, Z. Stuchlik, arXiv:1905.12452.  

\bibitem{r89} S. Hensh, Z. Stuchl\'{i}k, \emph{Eur. Phys. J. C} \textbf{79}, 834 (2019).

\bibitem{r90} V. Torres and E. Contreras, Eur. Phys. J. C, Eur. Phys. J. C {\bf 70}, 829 (2019).

\bibitem{r91}F. Linares and E. Contreras, Phys.\ Dark Univ. (Accepted), arXiv:1907.04892.

\bibitem{r92} S. Maurya, and F. Tello-Ortiz, arXiv:1907.13456

\bibitem{r93} R. Casadio, E. Contreras, J. Ovalle, A. Sotomayor, Z. Stuchlick,  	Eur. Phys. J. C {\bf 79}, 826 (2019)  

\bibitem{r94} A, Rinc\'on, L. Gabbanelli, E. Contreras and F. Tello-Ortiz, Eur. Phys. J. C {\bf 79}, 873 (2019)

\bibitem{roldao1}A. Fernandes-Silva, A.J. Ferreira-Martins, R. da Rocha,
Phys.Lett. B791 (2019) 323-330.

\bibitem{roldao2} R. da Rocha, \emph{Symmetry} \textbf{12} 508 (2020).

\bibitem{r95} J. Ovalle and R. Casadio, Beyond Einstein Gravity.
		The Minimal Geometric Deformation Approach in the Brane-World, Springer International Publishing (2020). DOI:10.1007/978-3-030-39493-6.
		
\bibitem{r96} E. Contreras, Eur. Phys. J. C {\bf 78}, 678 (2018).		



\bibitem{is} W. Israel, \emph{Nuovo Cim. B} \textbf{44}, 1 (1966). 

\bibitem{dar} G. Darmois, M\'emorial des Sciences Mathematiques (Gauthier-Villars, Paris, 1927), Fasc. 25 (1927).
\end{thebibliography}
\end{document}